\documentstyle[12pt]{article}
\catcode`@=11   
\def\NPB{{\it Nucl. Phys. }{\bf B}}
\def\PL{{\it Phys. Lett. }}
\def\PRL{{\it Phys. Rev. Lett. }}
\def\PRD{{\it Phys. Rev. }{\bf D}}
\def\CQG{{\it Class. Quantum Grav. }}

\def\IJMPA{{\it Int. J. Mod. Phys. }{\bf A}}

\def\ie{{\it i.e.}}

\newcommand{\eq}{\begin{equation}}
\newcommand{\en}{\end{equation}}
\newcommand{\eqn}{\begin{eqnarray}}
\newcommand{\enn}{\end{eqnarray}}
\newcommand{\nn}{\nonumber }
\newcommand{\beq}{\begin{equation}}
\newcommand{\eeq}{\end{equation}}

\let\a=\alpha

\let\b=\beta

\def\CP#1{\relax\ifmmode\IP^{#1}\else\IP$^{#1}$\fi}
\def\rd{{\rm d}}
\def\define{\buildrel{\rm def}\over=}
\let\f=\phi
\let\F=\Phi

\let\F=\Phi
\let\g=\gamma
\let\G=\Gamma

\def\inv#1{{\textstyle{1\over#1}}}
\def\IP{\relax\leavevmode{\rm I\kern-.18em P}}

\def\Ione{\relax\leavevmode{\rm 1\kern-3pt l}}
\let\q=\theta

\let\p=\pi
\let\vr=\varrho

\def\Seff{S_{\rm eff}}
\def\sgn{\mathop{\operator@font sign}\nolimits}
\let\t=\tau

\let\7=\widetilde

\let\vd=\partial
\def\vev#1{\left\langle#1\right\rangle}
\let\z=\zeta

\let\w=\omega

\def\frc#1#2{{\textstyle{#1\over#2}}}
\def\IR{\relax\leavevmode{\rm I\kern-.18em R}}
\def\ZZ{\relax\leavevmode
                   \ifmmode\mathchoice
                   {\hbox{\sf Z\kern-.4em Z}}
                   {\hbox{\sf Z\kern-.4em Z}}
                   {\lower.9pt\hbox{\scriptsize\sf Z\kern-.36em Z}}
                   {\lower1.2pt\hbox{\tiny\sf Z\kern-.36em Z}}
                    \else{\sf Z\kern-.4em Z}\fi}
\def\RR{\relax\leavevmode
                   \ifmmode\mathchoice
                   {\hbox{\sf R\kern-.4em R}}
                   {\hbox{\sf R\kern-.4em R}}
                   {\lower.9pt\hbox{\scriptsize\sf R\kern-.36em R}}
                   {\lower1.2pt\hbox{\tiny\sf R\kern-.36em R}}
                    \else{\sf R\kern-.4em R}\fi}

\def\resetby#1#2{\@addtoreset{#2}{#1}}
\def\seceq{\@addtoreset{equation}{section}
             \def\theequation{\thesection.\arabic{equation}}} 

\def\Label#1{\label{#1}%
         \smash{\hbox to0pt{\raise1ex\hbox{\tiny[#1]}\hss}}}
\def\noLabels{\let\Label=\label}

\thicklines     

\input epsf.tex
\reversemarginpar
\noLabels 
\catcode`@=12                   
\begin{document}

\begin{titlepage}
\begin{flushright}
CITUSC/01-009\\
hep-th/0104057\\
\end{flushright}

\begin{center}

{\large\bf\uppercase{
              Localized Gravity and Large Hierarchy
              from String Theory ?}}\\[10mm]
{\bf P. Berglund\footnote{e-mail: berglund@citusc.usc.edu} } \\[1mm]
              CIT-USC Center for Theoretical Physics\\
             Department of Physics and Astronomy\\
             University of Southern California\\
             Los Angeles, CA 90089-0484\\[5mm]
{\bf T. H\"{u}bsch\footnote{e-mail: thubsch@howard.edu}%
             $^,$\footnote{On leave from the ``Rudjer Bo\v skovi\'c''
             Institute, Zagreb, Croatia.} } \\[1mm]
             Department of Physics and Astronomy\\
             Howard University\\
             Washington, DC 20059\\[5mm]
{\bf D. Minic\footnote{e-mail: minic@citusc.usc.edu} } \\[1mm]
             CIT-USC Center for Theoretical Physics\\
             Department of Physics and Astronomy\\
             University of Southern California\\
             Los Angeles, CA 90089-0484\\[10mm]

{\bf ABSTRACT}\\[3mm]
\parbox{4.8in}{
Within a $D$-dimensional superstring spacetime, we construct a
non-supersymmetric brane-world with localized gravity and large
hierarchy between
the scale in the bulk, $M_D$, and the scale on the brane,
$M_{D-2}$.
  The localization of gravity and the large
hierarchy are both guaranteed by the presence of
non-trivial stringy moduli, such as the axion-dilaton system for the Type-IIB
string theory.
}

\end{center}

\end{titlepage}

In two recent papers~\cite{bhm1,bhm2},
we have studied non-supersymmetric solutions of string
theory which represent $p{+}1$-dimensional cosmic defects
embedded in a $D$-dimensional spacetime.
       These solutions explicitly realize a
possibility that our $3{+}1$-dimensional world can be identified with a
cosmic defect (brane) in a higher-dimensional theory, also studied
by many authors, for example~\cite{rubakov, visser, gell,
anton, georgia, horava, braneTH, savas, branew1, branew2, branew3,
rs1, rs2, cvetic, verlinde, greene, mayr}.
       The cosmic brane models of Refs.~\cite{bhm1,bhm2} feature a
large hierarchy between the $D$- and $(D{-}2)$-dimensional scales
induced from non-trivial warp factors in the metric and naturally
generated by the string coupling of $O(1)$.
       The space-time metrics considered in these models have been
studied before in the literature from a more phenomenological
point of view~\cite{cohen, cp, RG, tony}.
However, none of the solutions presented in the papers listed above
combine large hierarchy and localized gravity.

In this letter, we construct classical string theory
solutions by joining the two solutions found in Ref.~\cite{bhm2}
into a brane-world model with {\it both\/} naturally
localized gravity {\it and\/} large hierarchy. The large
hierarchy in our stringy brane-worlds is inherited from the general
solutions discussed in Ref.~\cite{bhm2}, and is caused by the dynamics of
the nontrivial stringy moduli. This large hierarchy is directly related
to the exponentially large volume of the transverse space of our solution
and can be therefore understood along the lines of \cite{savas}.
   Unlike the solutions of \cite{bhm1,bhm2}, in which gravity comes from the
usual Kaluza-Klein mechanism, the stringy brane-world models described in
this letter possess localized gravity due to the existence of a
normalizable zero mode, guaranteed by the presence of nontrivial moduli.
  Thus, it is the {\it nontrivial\/} stringy moduli which induce the
{\it simultaneous\/} appearance of localized gravity {\it and\/}
large hierarchy
on the {\it same\/} world-brane.

The general framework of this letter
is that of a higher-dimensional string theory compactified on a
Calabi-Yau (complex) $n$-fold, some moduli ($\f^\a$) of which are
allowed to vary over (the `transversal') part of the non-compact space.
Following Ref.~\cite{bhm1,bhm2}, the effective action describing the
coupling of the moduli to gravity of the observable spacetime
can be derived by dimensionally reducing the higher dimensional
Einstein-Hilbert action~\cite{vafa,gh}.
Thus, the relevant part of the low-energy
effective $D$-dimensional action of the moduli, $\f^\a$, of the Calabi-Yau
$n$-fold coupled to gravity reads
\begin{equation}
      \Seff = S_0 + \Seff^b
      = {1\over2\kappa^2}\int\rd^D x \sqrt{-g} ( R
         - {\cal G}_{\a \bar{\beta}}g^{\mu \nu}
           \vd_{\mu} \f^\a \vd_{\nu} \f^{\bar{\beta}}
                +...) + \Seff^b~.
\Label{e:effaction}
\end{equation}
Here $\mu,\nu=0,{\cdots},D-1$, $2\kappa^2=16\p G^D_N$, where $G_N^D$
is the $D$-dimensional Newton constant, and ${\cal G}_{\a \bar{\beta}}$ is the
metric on ${\cal M}_{\f^\a}$, the space of moduli. We neglect higher
derivative terms in this effective action and set the other fields in the
theory to zero as in~\cite{vafa}. We also restrict the moduli to depend on
$x_i$,
$i{=}D{-}2,D{-}1$, so that $\vd_a \f{=}0$,
$a{=}0,{\cdots},D{-}3$. $\Seff^b$ is a purely $D-1$-dimensional
effective action describing the brane world implied by the explicit
form of the solution described below. The equations of motion for the
moduli are
\begin{equation}
            g^{ij} \Big(\nabla_{i} \nabla_{j} \f^\a
            + \Gamma^\a_{\beta \gamma} (\f, \bar{\f}) \vd _{i}
            \f^{\beta} \vd_{j} \f^{\gamma} \Big) = 0~,\Label{e:scalar}
\end{equation}
where $\Gamma^\a_{\beta \gamma}$ is the connection on ${\cal M}_{\f^\a}$.
Note that $\Seff^b$ does not depend on the moduli. The
Einstein equations are
\begin{equation}
R_{\mu\nu} - \inv2 g_{\mu\nu} R
            = T_{\mu\nu} (\f,\bar{\f}) +
T_{\mu \nu}^b~,
\Label{e:einstein}
\end{equation}
where the energy-momentum tensor of the moduli is
\begin{equation}
            T_{\mu \nu}
            = {\cal G}_{\a \bar{\beta}} \Big(\vd_{\mu} \f^\a\vd_{\nu}
             \f^{\bar{\beta}} - \inv2 g_{\mu \nu}\, g^{\rho\sigma}
             \vd_{\rho} \f^\a \vd_{\sigma} \f^{\bar{\beta}}\Big)~
\end{equation}
and $T_{\mu \nu}^b$ is a delta-function source, as shown below.

We consider a single modulus scenario, $\f^\a{=}\tau$, and
in particular the axion-dilaton system, $\tau={\rm a} {+} ie^{-\F}$,
of the $D{=}10$ Type~IIB string theory, in which case $S_0$ in
(\ref{e:effaction}) describes the bosonic degrees of freedom of Type-IIB
supergravity.
     The explicit solutions for $\t$ exhibit non-trivial $SL(2,\ZZ)$
monodromy~\cite{bhm1,bhm2}, which ensures this to be a stringy rather
than (merely) a supergravity vacuum, and ${\cal
G}_{\tau\bar\tau}=-(\tau-\bar \tau)^{-2}$. With a phenomenologically
interesting
$K3$ compactification of the $D{=}10$ solution in mind (upon which the metric
receives $\alpha'$ corrections), we continue with
the general $D$-dimensional setting.
The metric that interpolates between the two solutions of
Ref.~\cite{bhm2}, with $z=\log(r)$, is:
\begin{eqnarray}
      \rd s^2 &=& A(z) \eta_{ab}\rd x^a \rd x^b
                + B(z) \rd z^2 + B(z) l^2 \rd\q^2~, \Label{e:Metric} \\[1mm]
      A(z) &=& Z^{{2\over D-2}}~,\qquad Z(z)\define 1+a_0 |z|~,
                \Label{e:A}\\[1mm]
      B(z) &=& Z^{-{D-3\over D-2}}\, e^{\frac{\xi}{a_0}[\b-Z^2]}~,
                \Label{e:B}
\end{eqnarray}
Here $\xi$, $a_0$ and $\beta$ are free parameters and $l$ sets the length
scale~\cite{bhm2}, which we will take to be of $O(M_D^{-1})$.
The dependence on $|z|$ (in place of just $z$ in
Refs.~\cite{bhm1,bhm2}) induces the $\delta$-function terms in
Eq.~(\ref{e:einstein})
(with $\vr\define\frc{D{-}3}{D{-}2}-2\frc{\xi}{a_0}$)
\begin{eqnarray}
   -\eta_{ab}\,l^{-2}\Big[a_0\xi\sgn^2(z)\, Z^{D-1\over D-2}\,
     e^{-{\xi\over a_0}(\b-Z^2)}\,
                  -e^{-{\xi\over a_0}(\b-1)}\,a_0\,\vr\,\delta(z)\Big]
   &=& T_{ab}+T^b_{ab}~,    \Label{e:Tab}\\
   -a_0\xi\,\sgn^2(z) &=& T_{zz}+T^b_{zz}~, \Label{e:Tzz}\\
   +a_0\xi\,\sgn^2(z)+2a_0\,\delta(z) &=& T_{\q\q}+T^b_{\q\q}~.
   \Label{e:Tqq}
\end{eqnarray}
Hereafter, we refer to the brane at $z=0$ as the {\it brane-world}:
there, $\sgn^2(z)=0$ and so $T_{\mu\nu}=0$ also. On the other hand,
the $\delta$-function terms in the left-hand side of the Einstein
equations~(\ref{e:einstein}) are now non-zero and read
\begin{equation}
    T_{\mu \nu}^b~=~l^{-2}\,\mbox{diag}
   \Big[-a_0\vr e^{-{\xi\over a_0}(\b-1)},
         a_0\vr e^{-{\xi\over a_0}(\b-1)},\ldots,
         a_0\vr e^{-{\xi\over a_0}(\b-1)},
         0, 2a_0\Big]\,\delta(z)~.
\end{equation}
  Since $\t$ depends on
$\w^2\equiv 8a_0\xi\geq0$~\cite{bhm1,bhm2}, we see that $\vr\geq0$ and
$\sgn(T_{00}^b)=-\sgn(a_0)$ for
  $|\xi|\leq|\xi_c|\define|{2(D-3)\over D-2}a_0|$.
     In particular, $T_{00}^b\geq0$ in the $a_0\leq0$ case, when $z$ is
restricted between the naked (null) singularities at $z=\pm1/a_0$.
When $\xi=0$ this form of the stress tensor is similar to that of spatial
domain walls~\cite{vilenkin,ipser}, in which the surface energy density,
$\sigma$, is equal to the surface tension, $-p$, where $p$ is the
pressure along the domain wall. In our case, however
$|T^b_{\q\q}|> T^b_{00}$.
 From this it follows
that the weak energy condition holds
except for
$T_{\q\q}^b$, \ie, $T_{\mu\nu}^b\z^\mu\z^\nu<0$ only for the null
vector $\zeta^\mu=(1,0,\cdots,0,\sqrt{A/B})$. (For a related discussion
of this feature of co-dimension two solutions consult for example
\cite{tinzul}.)
Still, we will assume that it is possible to associate an effective action
for the source at $z=0$,
\begin{equation}
   S^b_{\rm eff} = \int\rd^{D-2}x\,\rd
z\,\rd\q~\sqrt{-g}\,\delta(z)\,
                   \lambda\, {\cal L}^b~,\Label{e:Sbeff}
\end{equation}
Equating the $T^b_{\mu\nu}$ calculated from~(\ref{e:Sbeff})
with the $\delta$-function contribution of the Einstein equations from
Eq.~(\ref{e:Tab}--\ref{e:Tqq}), we obtain that
\begin{equation}
   \lambda\sim -a_0 l^{-2}e^{-{\xi\over a_0}(\beta-1)}~\qquad
   |\xi|\ll |\xi_c|~. \Label{e:V0}
\end{equation}
Note that the vacuum energy which couples to gravity is $\lambda\, {\cal
L}^b=-\lambda$.
Analogous results hold also in the $a_0>0,\xi>\xi_c$ case. However, now
$T_{\mu\nu}^b\z^\mu\z^\nu>0$ for {\it all} null vectors.

     We now demonstrate that, for a suitable choice of parameters, {\it both\/}
large hierarchy {\it and\/} localized gravity are realized in these
brane world solutions!

The large hierarchy between the $(D{-}2)$- and $D$-dimensional Planck
scales is the same as in Refs.~\cite{bhm1,bhm2}:
\begin{eqnarray}
      M_{D-2}^{D-4}&=&M_D^{D-2} \int_{M_\perp} \rd v\,\rd\q~\psi_0^2(v)
\Label{e:normalization}\\
&=&M_D^{D-2}{2\p l^2\over|a_0|}e^{\frac{\b\xi}{a_0}}
          \Big(\frac{a_0}{\xi}\Big)^{D-3\over2(D-2)}
          \cases{\Big[\G\Big(\frc{D-3}{2(D-2)}\Big)
                 -\g\Big(\frc{D-3}{2(D-2)};{\xi\over a_0}\Big)\Big]
          & for $a_0>0$,\cr
          \noalign{\vglue1mm}
          \g\Big(\frc{D-3}{2(D-2)};{\xi\over a_0}\Big)
          & for $a_0<0$.\cr
          }\qquad \Label{e:scaling}
\end{eqnarray}
where $M_\perp$ denotes the hyperbolic transverse
space~\cite{bhm2}.
     Note that the large hierarchy is controlled by the product of
$\b$ and the ratio ${\xi\over a_0}>0$, where the positivity of the latter
is due to the presence of the non-trivial stringy moduli. It is therefore
possible to choose
${\xi\over a_0} \b$, so as to have a
large hierarchy between $M_D$ and $M_{D-2}$.

Following the discussion of Randall
and Sundrum~\cite{rs1} we compute the coupling of gravity to the
fields on the brane. Writing,
$\bar g_{\mu\nu}\define g_{\mu\nu}|_{z=0}$ for the metric on the brane
world, there is a non-trivial contribution from $\sqrt{-g}$, \ie,
$\sqrt{-g}|_{z=0}=\sqrt{-\bar g}\, l^2e^{(\beta-1)\xi/a_0}$.
Hence,~(\ref{e:Sbeff}) becomes
\begin{equation}
   S^b_{\rm eff} \sim -a_0  \int\rd^{D-2}x\rd\q
~\sqrt{-\bar g}\,{\cal L}^b~,\Label{e:Sbeffnew}
\end{equation}
where we have taken into account the tension for the brane world
according to Eq.~(\ref{e:V0}). Thus, in contrast to the scenario
in~\cite{rs1} there is no need to rescale the fields, masses, couplings
and vevs in ${\cal L}^b$, and they retain their fundamental,
$D$-dimensional value, $O(M_D)$.
  Also, using Eq.~(\ref{e:Metric}), the kinetic terms of a
typical field, $\Psi$, expand
\begin{equation}
   |\vd_\mu \Psi|^2 = |\vd_\| \Psi|^2+l^{-2}e^{-{\xi\over
a_0}(\beta-1)}|\vd_\perp
\Psi|~,
\end{equation}
so that the transverse excitations of $\Psi$ are exponentially
supressed.

To understand the localization of gravity, we look at small gravitational
fluctuations $\delta\eta_{ab}=h_{ab}$
of the longitudinal part of the metric.
{}From the Einstein equations,
$h_{ab}$ satisfies a wave equation of the form \cite{csaki}:
\begin{equation}
     \Box h_{ab} = \frac{1}{\sqrt{-g}}\,\vd_{\mu}(\sqrt{-g}
                   g^{\mu\nu}\vd_{\nu} h_{ab}) = 0~. \Label{e:WEq}
\end{equation}
Following~\cite{cohen} we  change coordinates,
\begin{eqnarray}
      \rd v &=& l\,Z^{-\frac{(D-1)}{2(D-2)}}
            e^{\frac{\xi}{2a_0}[\b-Z^2]}\,\rd z~. \Label{e:dvdz} \\[1mm]
      \rd s^2 &=& A(v) \eta_{ab}\rd x^a \rd x^b
                + A(v) \rd v^2 + B(v) l^2 \rd\q^2~, \Label{e:vMetric}
\end{eqnarray}
The isometries of the metric dictate the following ansatz
\begin{equation}
      h_{ab}=\epsilon_{ab} e^{ip\cdot x} e^{in\q	} \frac{\phi}{\psi_0}~,
\end{equation}
where
\begin{equation}
        \psi_0\define\sqrt{A^{-1}\sqrt{-g}}
        =\sqrt{A^{\frac{D-3}{2}} B^{1\over 2}}
        =Z^{D-3\over4(D-2)}
         e^{\frac{\xi}{4a_0}[\b-Z^2]}~.
\end{equation}
With these variables~\cite{bhm1,bhm2, cohen}, Eq.~(\ref{e:WEq})
becomes a Schr\"{o}dinger-like equation:
\begin{equation}
         -\phi^{''} + \Big(\frac{\psi_0^{''}}{\psi_0} + \frac{A}{B}n^2\Big)\phi
          = m^2\phi~. \Label{e:Schrodinger}
\end{equation}
For simplicity, we choose $n=0$.
Integrating Eq.~(\ref{e:dvdz}) gives
\begin{eqnarray}
        v-v_0&=&\sgn(z)v_*\Big[\g_0^{-1}
         \g\Big(\frc{D-3}{4(D-2)};\frc{\xi Z^2}{2a_0}\Big)
          -1\Big]~, \nn\\[2mm]
        v_*&=&\frc{l}{2a_0}\,e^{\b\xi\over 2a_0}
               \Big(\frc{2a_0}{\xi}\Big)^{{D-3\over4(D-2)}}\,\g_0~,\qquad
        \g_0= \g\Big(\frc{D-3}{4(D-2)};\frc{\xi}{2a_0}\Big)~.
        \Label{e:VzChng}
\end{eqnarray}
The change of variables $z{\to}v$ is single-valued, continuous
and smooth across $z{=}0$, and $\sgn(z)=\sgn(v{-}v_0)$. Owing to the
dependence on $\g(a;x)$, Eq.~(\ref{e:VzChng}) is not {\it explicitly\/}
invertible, obstructing the evaluation of $\psi_0''/\psi_0$ in
Eq.~(\ref{e:Schrodinger}).
     However, in the $\xi\to0$ limit:
\begin{equation}
       \7v-v_0~=~\sgn(z)\,\7v_*\Big[Z^{D-3\over2(D-2)}-1\Big]~,
       \qquad \7v_*\define{2(D-2)\over D-3}{l\over a_0}~,
       \Label{e:vzChng}
\end{equation}
which is easy to invert {\it explicitly\/}. Hereafter, we set $v_0{=}0$.

Following the procedure described in the appendix of Ref.~\cite{bhm1},
the zero-mode wave function can be expressed, in the simpler $\xi{=}0$
case, in terms of $\7v$:
\begin{equation}
        \7\psi_0(\7v) = \sqrt{(1+|\7v|/\7v_*)}~.
\Label{e:psi0xi=0}
\end{equation}
The approximate effective potential
\begin{equation}
        \7V(\7v)\define\frac{\7\psi_0''}{\7\psi_0}
        =-\frac{\sgn^2(\7v)}{4(|\7v_*|-|\7v|)^2}
         -\frac{1}{|\7v_*|}\delta(\7v) \Label{e:TwiSchEq}
\end{equation}
exhibits a $\delta$-function well, which will remain true for small
enough $|\xi/a_0|{\neq}0$. By systematically expanding in $\xi$,
$\7V(\7v)$ may be turned into the true effective potential, $V(v)$. In
particular, for large enough $|\xi|$, the sign of the coefficient in front
of the $\delta$-function changes, and we must then choose $v_*>0$
for a $\delta$-function well. Here, we focus on the small-$|\xi|$
approximation~(\ref{e:TwiSchEq}), drop the tildes and defer the analysis of
the clearly
technically involved large-$|\xi|$ case for a more detailed
effort~\cite{bhm4}. (Equivalently we could consider the case in
which $a_0,\xi>0$ and expand around $v_\infty=v(|z|=\infty)$. One can
show that the result is exactly the same as in the situation considered
here~\cite{bhm4}.)

       The Schr\"{o}dinger equation can now be written as
\begin{equation}
        - \bigg[\frac{\rd^2}{\rd^2v}
               +\frac{\sgn^2(v)}{4(|v_*|-|v|)^2}
               +\frac{1}{|v_*|}\delta(v)
               \bigg]\phi_m = m^2 \phi_m~.
\end{equation}
Away from $v{=}0$, the general solution of this Schr\"{o}dinger equation is
\begin{equation}
\phi_m = a_m \sqrt{|v_*|-|v|}\,J_0\Big(m(|v_*|-|v|)\Big) +
                b_m \sqrt{|v_*| - |v|}\,Y_0\Big(m(|v_*|-|v|)\Big)~.
                \Label{e:Bessel}
\end{equation}
At $v{=}0$, the $\delta$-function implies the matching condition
\begin{equation}
         \bigg[-2\frac{d\phi_m}{d v}-\frac{1}{|v_*|}\phi_m\bigg]_{v=0}=0~.
\Label{e:delta}
\end{equation}
Evaluating Eq.~(\ref{e:Bessel}) for small values of $m$,
\begin{equation}
         \phi_m \sim a_m \sqrt{|v_*| - |v|}\,1 +
          b_m \sqrt{|v_*| - |v|}\,2\,\log\Big(\frc{1}{2}m(|v_*|-|v|)\Big)~,
\end{equation}
it is clear that Eq.~(\ref{e:delta}) is satisfied only if $b_m=0$.

     We are left to determine $a_m$ such that the normalization integral of
$\phi_m$ is $m$-independent
\begin{equation}
        \vev{\phi_m|\phi_m}
        =a_m^2\int_{-v_*}^{v_*} \rd v~(|v_*| - |v|) J^2_0(m(|v_*|-|v|)).
\Label{e:phimnorm}
\end{equation}
(Similar results of Eqs.~(\ref{e:psi0xi=0}-\ref{e:phimnorm}) hold for
$a_0,\xi>0$.) Note that this integral can be computed exactly, even
though the exact result is not relevant for the following discussion.
We now argue that this normalization integral is dominated by the plane
wave approximation, \ie,
\begin{equation}
        J_{0}^2(m|v_*|) \sim \frac{\cos^2(m|v_*| -\pi/4)}{m|v_*|},~\qquad
        m|v_*|\gg1~.
\Label{e:J0planewave}
\end{equation}
Strictly speaking, the plane wave approximation breaks down in the $\xi =0$
limit for the following reason:
The argument of the Bessel function solutions is
$m(|v_*|-|v|)$. If we assume that
$m\ll1/l$, since $l$ is the fundamental length scale in the problem (and
the higher dimensional scale), then $m(|v_*|-|v|)\ll|a_0|^{-1}$.
Therefore, it is not possible to approximate the
solution in terms of plane waves.

Fortunately, this problem is resolved by the presence of non-trivial
moduli, \ie, by $\xi\neq0$, which is necessary for the large
hierarchy. In this case we obtain the same general
differential equation, except that an explicit expression for $V(v)$ is
lacking. However, we may replace, in Eq.~(\ref{e:B}),
$e^{\frac{\xi}{a_0}[\beta-Z^2]}\to e^{\frac{\beta \xi}{a_0}}$, which is a
good approximation near the naked singularities, at
$|z|=|a_0|^{-1}$. Nevertheless, the result is surprisingly accurate.
       The zero-mode wave function can be
expressed in terms of $v$,
just as before. Note that $v_*$ now depends on $\xi$.
The normalization integral for $\psi_0$ becomes,
\begin{equation}
    \vev{\psi_0|\psi_0}= 2\p\int \rd v~|\psi_0|^2
    = \frac{2(D-2)}{D-3}\,\frac{2\p l^2}{|a_0|}\,e^{{\b|\xi|\over|a_0|}}~.
\end{equation}
When we compare this expression with the  exact result~(\ref{e:scaling}),
the only discrepancy occurs in the power of $a_0/\xi$ and the
overall $O(1)$ numerical factor.
Similar arguments apply for the $\phi_m$ when $\xi\neq 0$. For
the normalization $\vev{\phi_m|\phi_m}$ we get,
\begin{equation}
        \vev{\phi_m|\phi_m}_{\xi\neq 0} =e^{{\b|\xi|\over|a_0|}}
         \vev{\phi_m|\phi_m}_{\xi=0}.
\Label{e:vevphim}
\end{equation}
The plane wave approximation is valid for $\xi \neq 0$ because
$v_*\sim l/|a_0| \exp({{\b|\xi|\over|a_0|}})$ and hence
when $m$ is large $m v_*\gg1$. Recall that while $m$ can be large, it
is limited by $l^{-1}$.
Thus, we can compute
$\vev{\phi_m|\phi_m}$ by looking at large
$m v_*$ for which the Bessel function, $J_0$, looks like a plane
wave~(\ref{e:J0planewave}). This means that
$\vev{\phi_m|\phi_m} \sim a_m^2 m^{-1} v_*$.
Thus, we
have to choose
$a_m\sim\sqrt{m}$. Since $v_* \sim l$, then
$\phi_m \sim\sqrt{m l}$.
It is now clear that $\psi_0\neq\lim_{m\to0}\phi_m$, \ie, the presence of
non-trivial stringy moduli guarantees the existence of the isolated zero
mode, and hence localized gravity at $z{=}0$.

With these facts, the Newton potential has the following form:
\begin{equation}
        U(r) \sim M_D^{-(D-2)} \frac{M_1 M_2}{r}
          \frac{\psi_0^2(0)}{ \vev{\psi_0|\psi_0}}
        + \sum_mM_D^{-(D-2)} \frac{M_1 M_2}{r} e^{-mr}
          \frac{\phi_m^2(0)}{ \vev{\phi_m|\phi_m}}~. \Label{e:Newt}
\end{equation}
The Newton constant is determined from the first term:
\begin{equation}
         G_N \sim M_D^{-(D-2)} \frac{\psi_0^2(0)}{ \vev{\psi_0|\psi_0}}~.
\end{equation}
Dimensional analysis implies that the correction to Newton's potential is
\begin{equation}
        \triangle U(r) \sim G_N \frac{M_1M_2}{r}  \int\rd(ml)\> (ml)^2 N
e^{-mr}~,
\end{equation}
where the integral approximates the large sum over dense modes in
Eq.~(\ref{e:Newt}). Here $(ml)^2$ comes from the integration measure and
$\phi_m^2(0)$, and
\begin{equation}
        N= \frac{ \vev{\psi_0|\psi_0}}
           { \vev{\phi_m|\phi_m}}\frac{(\phi^2_m(0)/m)}{\psi_0^2(0)}~.
\Label{e:N}
\end{equation}
It can be argued that to leading
order in $m$, $N$ does not depend on $m$ and is of $O(1)$.
The $m$ independence follows from the plane wave approximation, which we have
argued is valid in the $\xi \neq 0$ case.
Strictly speaking, higher order correction in $m$ give rise to
higher order corrections in $1/r$  to the Newtonian potential, which are
even more suppressed.

Given the form of the normalization $ \vev{\phi_m|\phi_m}$ from
(\ref{e:vevphim})
it is possible to compute $N$ by
looking at the $\xi=0$ case as the factor $\exp({{\b|\xi|\over|a_0|}})$ gets
canceled between the numerator and denominator in~(\ref{e:N}). Thus, we get
that $N\sim O(1)$, and so
\begin{equation}
{\triangle U(r)} \sim G_N \frac{M_1M_2}{r} \frac{l^3}{r^3}.
\end{equation}
Note that $\triangle U(r)$ does not depend on $|a_0|$
since $N\sim O(1)$ and
the $|a_0|$ factors cancel
between $\vev{\phi_m|\phi_m}$ and $\vev{\psi_0|\psi_0}$.
Also, ${\triangle U(r)}$ is a very small correction if, for example,
$M_D\sim TeV$, since
$l\sim (M_D)^{-1}$. The
Newton potential has only been checked
down to $r_e\sim 1\, \hbox{mm}\sim 10^{-12} GeV^{-1}$, so that
$l/r < l/r_e \sim 10^{-15}$.

In conclusion we discuss some stringy aspects of the brane world
described in this letter:
The
general framework of our analysis
is that of a higher-dimensional string theory compactified on a
Calabi-Yau (complex) $n$-fold, some moduli ($\f^\a$) of which are
allowed to vary over (the `transversal') part of the non-compact space.
In
the uncompactified Type~IIB theory, the r\^ole of space-dependent moduli
is played by the axion-dilaton system with $SL(2,\ZZ)$ monodromy
properties very much like in Vafa's description of F-theory~\cite{FTh}.
It is these non-holomorphic and non-trivial moduli that are responsible
for the localization of gravity and large hierarchy (the latter
being understood along the lines
of the scenarios with large extra dimensions \cite{savas}) in the brane
world solution discussed in this letter. Generically, our solutions have
naked singularities (the resolution of which was discussed in
\cite{bhm2,bhm3}).
Consequently, our solutions do not saturate the BPS bound in general,
supersymmetry being
broken in the presence of the naked singularities, and can be thought of as
a particular example of warped Kaluza-Klein compactifications in string
theory. However, as shown in \cite{bhm2}, there exists a limit in which our
solution formally becomes supersymmetric. The hyperbolic transverse space
degenerates in this case into a cylinder. This supersymmetric limit (as
discussed in~\cite{bhm2}) describes a collection of strongly interacting
$D7$-branes.
   Upon a $K3$ compactification, the product of this $K3$ and the transverse
cylinder may also be identified with a $K3$ fibered Calabi-Yau 3-fold, in
the limit where the base $\IP^1$ degenerates into the transverse cylinder.
The metric in such a limit may be analyzed along the lines in
Ref.~\cite{cand}.

Now, in analogy with Verlinde's discussion
of the origin of the Randall-Sundrum solution from Calabi-Yau
compactifications~\cite{verlinde},
we might envision a situation in which, after supersymmetry breaking,
the form factors in the longitudinal part of the metric and the
2-dimensional part of the
Calabi-Yau metric
may be well approximated by
the warp factors describing our solution.
The difficulty here, as opposed to Verlinde's discussion, is that
our solution is explicitly non-supersymmetric. In Verlinde's discussion the
Randall-Sundrum brane world is made of $D3$-branes. In our case we
unfortunately
do not understand the nature of the non-supersymmetric $3$-brane obtained
after wrapping a distribution of $7$-branes on a K3.
Thus the discussion becomes even more qualitative.

{\bf Acknowledgments:}
We thank V.~Balasubramanian, J.~de~Boer, C.~Johnson, M.~Luty, P.~Mayr
and J.~Polchinski for useful discussions.
P.~B. would like to thank the theory group at SLAC for its
hospitality in the final stages of this project.
            The work of P.~B.\ was supported in part by  the US
Department of Energy under grant number DE-FG03-84ER40168.
            T.~H.\ wishes to thank the Caltech-USC Center for Theoretical
Physics for its hospitality, and the US
Department of Energy for their generous support under grant number
DE-FG02-94ER-40854.
            The work of D.~M.\ was supported in part by the US
Department of Energy under grant number DE-FG03-84ER40168.

\end{document}